\documentclass[aps,prl, twocolumn] {revtex4}
\usepackage{epsfig}

\begin{document}
\headsep 2.5cm

\title{Beyond real space super cell approximation, corrections  to the real space cluster approximation . }
\author{Rostam Moradian$^{1,2}$, Sina Moradian$^3$}

\affiliation{$^{1}$Physics Department, Faculty of Science, Razi
University, Kermanshah, Iran\\
$^{2}$Nano Science and Nano Technology Research Center, Razi University, Kermanshah, Iran\\
$^3$Department of Electrical and Computer Engineering, University of Central Florida, Orlando, Florida, USA.}


\begin{abstract}
Motion of a single electron in a disordered alloy and or interacting electrons systems such as magnetic materials, strongly correlated systems and superconductors is replaced by motion of that in an effective medium which is denoted by self-energy. The study of disordered alloy and interacting electrons systems based on single electron motion is an old challenge and an important problem in condensed matter physics. In this paper we introduce a real space approximation beyond super cell approximation for the study of these systems to capture multi-site effects. Average disordered alloy or interacting system is replaced by a self-energy, $\Sigma(i,j,E)$. We divided self-energy in q-space $\Sigma({\bf q}; E)=\frac{1}{N}\sum_{ij}e^{i{\bf q}.{\bf r}_{ij}}\Sigma(i,j; E)$ into two parts $\Sigma({\bf q}; E)=\frac{1}{N_{c}}\sum_{IJ\in\; \mbox{\tiny same cluster}}e^{i{\bf q}.{\bf r}_{IJ}}\Sigma(I,J; E)+\frac{1}{N}\sum_{ij\notin \:\mbox{\tiny same cluster}}e^{i{\bf q}.{\bf r}_{IJ}}\Sigma(I,J,E)$  where $\{Lc_{1}, Lc_{2},Lc_{3}\}$ are dimensions of the super cell. We show that neglecting the second term of q-space self-energy leads to super cell approximation $e^{iq_{j} Lc_{j}}=1$, hence $ q_{j}$ determined by $ q_{j} Lc_{j}=2\pi n_{j}$. Then we kept this correction in the second step to add self energies of sites in different super cells which leads to fully q-dependent self energy in the first Brillouin zone (FBZ). Our self-energy in FBZ is casual, fully q-dependent, and continuous with respect to ${\bf q}$. It recovers coherent potential approximation in the single site approximation and is exact when the number of sites in the super cell approaches to the total number of lattice sites.  We illustrate that this approximation undertakes electrons localization for one and two dimensional alloy systems which isn't observed by previous multi site approximations.
\end{abstract}

\flushbottom
\maketitle

%
%
\thispagestyle{empty}
\section{Introduction}
The treatment of disordered and interacting electron systems based on single electron motion in an effective medium is an important problem in many fields such as alloys, strongly correlated systems, magnetism and superconductivity in condensed matter physics. Coherent Potential Approximation (CPA) and Dynamical Mean Field Theory (DMFT) are single site approximations for calculating effective medium denoted by self-energy. In these approximations muti-site effects is neglected.  Metzner and Vollhardt\cite{Metzner} and Muler-Hartmann\cite{Muller-Hartmann} found that in the limit of infinite dimensions both single site approximations coherent potential approximation (CPA) \cite{Soven67} for disordered system and dynamical mean field theory for interacting systems are exact\cite{Metzner, Muller-Hartmann}. This means self-energy for systems with high dimensions is k-independent. However, outside of systems with infinite dimensions especially in one and two dimensional systems self-energy is far from local which means it is k-dependent. To treat effective features of disorder systems, k-dependent relation of self-energy $\Sigma({\bf k}; E)$ must be identified. In lower approximations such as Born approximation, T-matrix approximation, and Coherent Potential Approximation (CPA) \cite{Soven67}, which are single site approximations, self-energy is k-independent $\Sigma({\bf k}; E)=\Sigma( E)$. In these approximations multi-site scattering is neglected which leads to overestimation band splitting and also losing short range effects. To add these effects cluster CPA with k-independent self-energy is used \cite{Gonis}. Dynamical Cluster Approximation (DCA)\cite{Hettler98, Jarrell01, Jarrell01-2} for systems with weak k-dependent self energies by considering  periodic boundary condition for both interacting electron and disorder systems introduced. Also cellular dynamical mean field theory (CDMFT) approximation with open boundary condition\cite{Kotliar} was introduced. In DCA, first Brillouin zone (FBZ) is divided to $N_{c}$ grain regions where self-energy inside of these grains is k-independent although they could be different. So their self-energy in the FBZ is not continues.  The wave vectors at center of these grains called cluster wave vectors and are denoted by $\{{\bf K}_{1}, ...,{\bf K}_{N_{c}}\}$. For disordered systems they claimed these cluster wave vectors, $\{{\bf K}_{n}\}$, corresponds to a $N_{c}$ real cluster sites \cite{Hettler98}. Although both cluster approximations DCA and CDMFT are successful in importing multi site effects but these methods have two major weakness, first their grain self energies are discontinuous, second at low dimension self energy is strongly k-dependent. In real space effective medium super-cell approximation (EMSCA)\cite{Moradian02, Moradian04} are used to approximate self-energy of interacting disordered systems.  We show that DCA could be a super cell approximation. Here in real space we first introduce super cell approximation by neglecting k-space self-energy contribution of all sites in different super cells. Then by keeping this contribution we go beyond super cell approximation. In our formalism, self  energy is k-dependent and continuously varying in FBZ. Our self-energy is more close to real self-energy. Hence the average Green function calculated with our method used for calculation of physical quantities is more close to real average Green function.

The organization of the paper is as follows.
In Sec. II the  model Hamiltonian and super cell approximation equations are presented.
Beyond super cell approximation equations derived and applied to a two dimensional alloy system in Sec. III. In this section we calculated and compared density of states for single site, super cell and beyond super cell approximations. Also electron localization in these approximations are discussed.
\section{Model Hamiltonian and self-energy in the super cell approximation}
The starting point is a  tight-binding model for a disorder alloy system which is given by,
\begin{eqnarray}
H&=&-\sum_{ij\sigma\sigma}t_{ij}c^{\dagger}_{i\sigma}c_{j\sigma}
\nonumber\\&+&\sum_{i\sigma} (\varepsilon_{i}-\mu)c^{\dagger}_{i\sigma}c_{i\sigma},
\label{eq:Hamiltonian}
\end{eqnarray}
where $c^{\dagger}_{i\sigma}$ ($c_{i\sigma}$) is the creation (annihilation) operator of an electron with spin $\sigma$ on lattice site $i$ and $\hat{n}_{i\sigma}=c^{\dagger}_{i\sigma}c_{i\sigma}$ is the number operator. $t^{\sigma\sigma}_{ij}$ are the hopping integrals between $i$ and $j$ lattice sites with spin $\sigma$ respectively. $\varepsilon_{i}$ is the random on-site energy and takes  $-\delta$ with probability  $1-c$ for the host sites and $\delta$ with probability $c$ for impurity sites and $\mu$ is the chemical potential.

The electron equation of motion for Hamiltonian, Eq.\ref{eq:Hamiltonian}, is given by,
\begin{eqnarray}
\sum_{l} \left(
       \begin{array}{c}
(E-\varepsilon_{i}+\mu)\delta_{il}-t_{il}\end{array}\right){ G}(l,j; E)=\delta_{ij}
\label{eq:equation of motion}
\end{eqnarray}
where $G(i,j)$ is the random single particle Green function. Relation between electron's random Green function matrix, ${\bf G}$, and average Green function matrix, ${\bf \bar G}$, is given by
\begin{equation}
 {\bf G}=\bf{\bar{G}}+\bf{\bar{G}}({\mbox{\boldmath$\varepsilon$}}-{\mbox{\boldmath$\Sigma$}}){\bf G}.
\label{eq:random average dyson equation}
\end{equation}

\begin{figure}[ht]
\centering
\includegraphics[width=6cm]{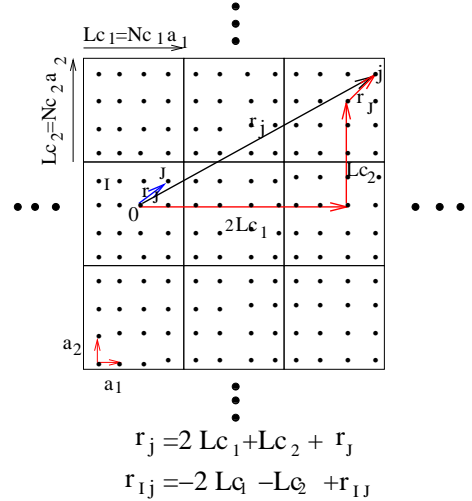}
\caption{(Color online) Show a two dimensional square lattice which is divided to similar super cells of $N_{c}=16$ with original lattice symmetry. The super cell vectors are ${\bf Lc}_{1}=4 a{\bf e}_{x}$ and ${\bf Lc}_{2}=4 a {\bf e}_{y}$ where $\{{\bf a}_{1}=a{\bf e}_{x}, {\bf a}_{2}=a{\bf e}_{y}\}$ and {\em  a} is lattice constant. }
 \label{figure:real-space-nlcpa}
\end{figure}

 Note that although \ref{eq:random average dyson equation} is exact, due to randomness no exact solutions exists. For calculation of self-energy, different single site approximations such as coherent potential approximation (CPA), T-matrix, and Born approximation are introduced. Although attempts have been made with DCA to include multi-site scattering, its coarse grained self energies are discontinuous therefore in k-space these attempts have been unsuccessful. A real space multi site  approximation which preserves continuity of k-space dependence of self-energy in the FBZ is not introduced. Here we implement a real space cluster approximation beyond super cell approximation in which not only includes multi site scattering but in the FBZ k-space dependence of self-energy varying continuously. Consider a lattice with dimensions $\{{\bf L}_{1}=N_{1}{\bf a}_{1}, {\bf L}_{2}=N_{2}{\bf a}_{2}, {\bf L}_{3}=N_{3}{\bf a}_{3}$ and sites number $N=N_{1}N_{2}N_{3}$ where ${\bf a}_{j}$ are lattice primitive vectors. Divide this lattice to super cells with dimensions $\{{\bf Lc}_{1}=N_{c1}{\bf a}_{1},{\bf Lc}_{2}=N_{c2}{\bf a}_{2},{\bf Lc}_{3}=N_{c3}{\bf a}_{3}\}$, original lattice symmetries and super cell lattice sites number $N_{c}=N_{c1}N_{c2}N_{c3}$. Position of $N_{c}$ sites in side of each cell denoted by capital letters $\{I\}$. Number of super cells is $\frac{N}{N_{c}}$. Since for alloy system at the band splitting regime for $c=0.5$ and average band filling ${\bar n}=1$ all sites with onsite energy $\delta$ are empty with sites with onsite energy $-\delta$ are filled by two electrons, just super cell with even sites number are acceptable. Fig.\ref{figure:real-space-nlcpa} shows this for a two dimensional square lattice with $N_{c}=16$. Note that impurity configuration of super cells are not same.

\begin{figure}[ht]
\centering
\includegraphics[width=7cm]{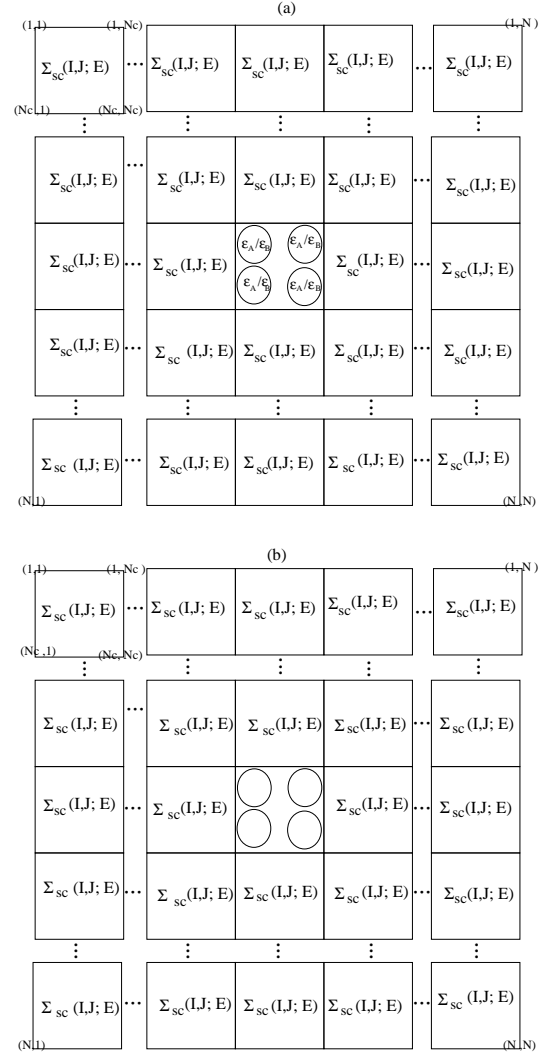}
\caption{(color online) (a ) shows an  $N_{c}=4$ impurity super cell in an effective super cell self-energy medium in the lattice sites space matrix. (b) a $N_{c}=4$ cavity super cell embedded in super cell self energies in real space lattice sites matrix. }  \label{figure:impurity-super-cell}
\end{figure}

Since real space self energies only depend on difference of two lattice sites positions $\Sigma(E;i,j)=\Sigma(i-j; E)$, self energies divided to two categories,  first self energies between intra sites of each super cell, $\Sigma(I-J; E)$, second self energies of one site inside of a super cell but another site belongs to another super cell $\Sigma(I-j; E)$ in which
\begin{eqnarray}
{\bf r}_{j}=l{\bf Lc}_{1}+m{\bf Lc}_{2}+n{\bf Lc}_{3} +{\bf r}_{J} .
\label{eq:vector}
\end{eqnarray}
where $l=\{0,1,...,\frac{N_{1}}{N_{c1}}-1\}$, $m=\{0,1,...,\frac{N_{2}}{N_{c2}}-1\}$, $n=\{0,1,...,\frac{N_{3}}{N_{c3}}-1\}$ are integer numbers. The exact q-space self-energy is
\begin{eqnarray}
 &\Sigma&({\bf q}; E)=\frac{1}{N_{c}}\sum_{IJ}  \Sigma(I, J; E)e^{i{\bf q}.{\bf r}_{IJ}}+\sum_{\bf q'} \Sigma({\bf q'}; E)\frac{1}{N_{c}N}\nonumber\\&\times&\sum_{IJ} e^{i({\bf q}-{\bf q'}).{\bf r}_{IJ}}\Pi^{3}_{j=1} \left(\frac{1-e^{-iN_{j}a_{j}(q_{j}-q'_{j})}}{1-e^{-iN_{cj}a_{j}(q_{j}-q'_{j})}}-1\right).
\label{eq:self energy-k2-4-0}
\end{eqnarray}
 Our first approximation for exact q-space self-energy Eq.\ref{eq:self energy-k2-4-0} is that, contribution of summation over all two lattice sites in different super cells become zero,
\begin{eqnarray}
 \left(\frac{1-e^{-iN_{j}a_{j}(q_{j}-q'_{j})}}{1-e^{-iN_{cj}a_{j}(q_{j}-q'_{j})}}-1\right)=0
\label{eq:super-cell-self energy-k2-4-1}
\end{eqnarray}
The Born von Karman periodic boundary condition\cite{Ashcroft87} imply that $e^{-iN_{j}a_{j}(q_{j}-q'_{j})}=1$, hence in Eq.\ref{eq:super-cell-self energy-k2-4-1} we have
\begin{eqnarray}
e^{-iN_{cj}a_{j}(q_{j}-q'_{j})}=1.
\label{eq:super-cell-self energy-k-condi}
\end{eqnarray}
From Eq.\ref{eq:super-cell-self energy-k-condi} we have
\begin{eqnarray}
   N_{cj}{\bf a}_{j}.{\bf q}_{j}=2\pi n_{j},\;\;j=1,2,3  .
\label{eq:math3}
\end{eqnarray}

where $\{n_{j}\}$ are integer numbers such that one of  ${\bf q}$ must be center of FBZ. Wave vectors ${\bf q}={\bf K}_{n}$ that satisfy Eq.\ref{eq:math3} are
\begin{eqnarray}
  {\bf K}_{n}={\bf K}_{m_{1}m_{2}m_{3}}=\sum^{3}_{i=1}\frac{m_{i}}{N_{ci}} {\bf b}_{i}  .
\label{eq:math4}
\end{eqnarray}
  where $\{{\bf b}_{1}, {\bf b}_{2}, {\bf b}_{3}\}$ are reciprocal lattice primitive vectors, $\{m_{1},\;m_{2},\;m_{3}\}$ are integer such that ${\bf K}_{n}$ remains in the FBZ. By substitution Eqs.\ref{eq:math4} and \ref{eq:super-cell-self energy-k-condi} in to Eq.\ref{eq:self energy-k2-4-0} we obtain
\begin{eqnarray}
 &\Sigma&({\bf K}_{n'}; E)=\frac{1}{N_{c}}\sum_{IJ}  \Sigma_{sc}(I, J; E)e^{i{\bf K}_{n'}.{\bf r}_{IJ}}
\label{eq:self energy-K_n'}
\end{eqnarray}
By times both sides of Eq.\ref{eq:self energy-K_n'} by $e^{-i{\bf K}_{n'}.{\bf r}_{I'J'}}$ and summation over ${\bf K}_{n'}$ and using this fact that
\begin{eqnarray}
 \frac{1}{N_{c}}\sum_{J}  e^{i({\bf K}_{n}-{\bf K}_{n'}).{\bf r}_{J}}=\delta_{{\bf K}_{n}{\bf K}_{n'}},\;\; \frac{1}{N_{c}}\sum_{{\bf K}_{n}}  e^{i{\bf K}_{n}.{\bf r}_{IJ}}=\delta_{IJ}\nonumber\\
\label{eq:self energy-K_n'f}
\end{eqnarray}
we have
\begin{eqnarray}
 \Sigma_{sc}(I, J; E)=\frac{1}{N_{c}}\sum_{{\bf K}_{n'}}  \Sigma({\bf K}_{n'}; E)e^{-i{\bf K}_{n'}.{\bf r}_{IJ}}.
\label{eq:self energy-I-J}
\end{eqnarray}
Eqs.\ref{eq:super-cell-self energy-k-condi} and \ref{eq:math4}
 imply that in this approximation $e^{-ilN_{c1}{\bf a}_{1}.{\bf K}_{n'}}=1$, $e^{-imN_{c2}{\bf a}_{2}.{\bf K}_{n,}}=1$, $e^{-inN_{c3}{\bf a}_{3}.{\bf K}_{n'}}=1$. By use of Eq.\ref{eq:self energy-I-J} and considering Eq.\ref{eq:vector}, real space self energies of two sites $I$ and $j$ in different super cells are  periodic with respect to super cells center vector position
\begin{eqnarray}
 \Sigma_{sc}(I,j; E)=\Sigma_{sc}(I,J; E).
\label{eq:self energy approximation1}
\end{eqnarray}
 Therefore in lattice sites, space self-energy matrix is constructed from just super cell self-energy matrices. This illustrated in Fig.\ref{figure:impurity-super-cell}.
Note that for $N_{c}=1$, ${\bf K}_{n}=0$ hence converts to CPA self-energy which is k-independent. For  $lim_{N_{c}\rightarrow N;\;I,J\rightarrow i,j}\;\frac{1}{N_{c}} \sum_{J}e^{i{\bf K}_{n}.{\bf r}_{IJ}} \Sigma(I,J;E)= \frac{1}{N} \sum_{ij} e^{i{\bf k}.{\bf r}_{ij}} \Sigma(i,j;E)= \Sigma({\bf k};E)$ it is exact k-space self-energy.

By taking impurity average over all random lattice sites except central super cell sites, Eq.\ref{eq:random average dyson equation} reduces to a $N_{c}\times N_{c}$ matrix of super cell impurity embedded is an effective medium of super cell self energies
\begin{equation}
 {\bf G}^{im}_{N_{c}\times N_{c}}={\bf\bar{G}}_{N_{c}\times N_{c}}+{\bf\bar{G}}_{N_{c}\times N_{c}}({\mbox{\boldmath$\varepsilon$}}-{\mbox{\boldmath$\Sigma$}}_{sc})_{N_{c}\times N_{c}} {\bf G}^{im}_{N_{c}\times N_{c}}.
\label{eq:imp-random average dyson equation}
\end{equation}
as illustrated in Fig.\ref{figure:impurity-super-cell} (a). Eq.\ref{eq:imp-random average dyson equation} can be written as
\begin{equation}
 {\bar{\bf G}}^{-1}_{sc}-{\mbox{\boldmath$\Sigma$}}_{sc}= {{\bf G}^{im}_{sc} }^{-1}-{\mbox{\boldmath$\varepsilon$}}={\mathcal G} .
\label{eq:cavity-imp-random average dyson equation}
\end{equation}
where ${\mathcal G}$ is called cavity super cell Green function as shown in \ref{figure:impurity-super-cell}(b).
Eq.\ref{eq:cavity-imp-random average dyson equation} separates to two following super cell Dysons like equations
\begin{eqnarray}
  G^{im}_{sc} (I,J; E)={\mathcal G}(I,J; E)+ \sum_{L}{\mathcal G}(I,L; E) \varepsilon_{L}G^{im}_{sc} (L,J; E).\nonumber\\
\label{eq:math5-1}
\end{eqnarray}
and
\begin{eqnarray}
  &{\bar G}_{sc} &(I,J; E)={\mathcal G}(I,J; E)+\nonumber\\ &\;&\sum_{LL'}{\mathcal G}(I,L; E)\Sigma_{sc} (L,L'; E){\bar G}(L',J; E).
\label{eq:math5-2}
\end{eqnarray}
The Fourier transform of real space super cell average Green function and cavity Green function to super cell wave vectors $\{{\bf K}_{n}\}$ and vice versa are
\begin{eqnarray}
 {\bar G}({\bf K}_{n}; E)&=& \frac{1}{N_{c}}\sum_{IJ}{\bar G}_{sc} (I,J; E)e^{-i{\bf K}_{n}.{\bf r}_{IJ}},\nonumber\\ {\bar G}_{sc} (I,J;E)&=& \frac{1}{N_{c}}\sum_{{\bf K}_{n}}{\bar G}({\bf K}_{n}; E)e^{i{\bf K}_{n}.{\bf r}_{IJ}}
\label{eq:math5-3}
\end{eqnarray}
\begin{eqnarray}
 {\mathcal G}({\bf K}_{n}; E)&=& \frac{1}{N_{c}}\sum_{IJ}{\mathcal G}(I,J; E)e^{-i{\bf K}_{n}.{\bf r}_{IJ}},\nonumber\\{\mathcal G}(I,J; E)&=& \frac{1}{N_{c}}\sum_{{\bf K}_{n}}{\mathcal G}({\bf K}_{n}; E)e^{i{\bf K}_{n}.{\bf r}_{IJ}}.
\label{eq:math5-4}
\end{eqnarray}
Substituting Eqs.\ref{eq:self energy-I-J} , \ref{eq:math5-3} and \ref{eq:math5-4} in Eq.\ref{eq:math5-2} and using Eq.\ref{eq:self energy-K_n'f} we have
\begin{eqnarray}
  {\bar G}({\bf K}_{n}; E)={\mathcal G}({\bf K}_{n}; E)+ {\mathcal G}({\bf K}_{n}; E) \Sigma({\bf K}_{n};E){\bar G}({\bf K}_{n}; E).
\label{eq:math5-7}
\end{eqnarray}

To calculate $\Sigma({\bf K}_{n}; E)$ the FBZ is divided into $N_{c}$ regions with FBZ symmetries and $\frac{N}{N_{c}}$ wave vectors where each of ${\bf K}_{n}$ are in the center of one of these grains .Inside each grain self-energy is k-independent therefore, it is grain CPA self-energy. At $lim_{N_{c}\rightarrow N}$ number of wave vectors $\{\bf k\}$ in each grain reduces to just one (${\bf K}_{n}={\bf k}$). The $n$th grain CPA average Green function is defined by
\begin{eqnarray}
  \bar{G}({\bf K}_{n}; E)=\frac{N_{c}}{N}\sum_{{\bf k}\in nth\; grain}\frac{1}{E-\epsilon_{\bf k}+\mu-\Sigma({\bf K}_{n}; E)}.
\label{eq:nth grain green}
\end{eqnarray}
and its real space Fourier transform is
\begin{eqnarray}
  \bar{G}(I,J; E)=\frac{1}{N_{c}}\sum_{{\bf K}_{n}}\bar{G}({\bf K}_{n}; E)e^{-i{\bf K}_{n}.{\bf r}_{IJ}} .
\label{eq:real space nth grain green}
\end{eqnarray}

Note that DCA\cite{Jarrell01-2} could be real space super cell approximation.

\section{beyond super cell approximation}
 To go beyond super cell approximation and add  self energies contribution of $i$ and $j$ which are not in the same super cell we use super cell approximation $\Sigma(I,J; E)\approx\Sigma_{sc}(I,J; E)=\frac{1}{N_{c}}\sum_{{\bf K}_{n}}\Sigma({\bf K}_{n}; E)e^{i{\bf K}_{n}.{\bf r}_{IJ}}
$ hence
\begin{eqnarray}
\sum_{IJ}  \Sigma(I,J; E)e^{i{\bf k}.{\bf r}_{IJ}}\approx \frac{1}{N_{c}}\sum_{IJ}\sum_{{\bf K}_{n}}\Sigma({\bf K}_{n}; E) e^{i({\bf k}-{\bf K}_{n}).{\bf r}_{IJ}}.\nonumber\\
\label{eq:byond-super-cell}
\end{eqnarray}
Note that beyond super cell approximation where $1<N_{c}<N$, for $q_{j}\neq K_{nj}$ we have $1-e^{-iN_{cj}a_{j}(q_{j}-q'_{j})}\neq 0$. By inserting Eq.\ref{eq:byond-super-cell} in to Eq.\ref{eq:self energy-k2-4-0} we have
\begin{eqnarray}
\Sigma({\bf k}; E)&=&\frac{1}{N^{2}_{c}}\sum_{IJ}\sum_{{\bf K}_{n}}\Sigma({\bf K}_{n},E) e^{i({\bf k}-{\bf K}_{n}).{\bf r}_{IJ}}-\nonumber\\&\;&\sum_{\bf q'} \Sigma({\bf q'}; E)\frac{1}{N_{c}N}\sum_{IJ} e^{i({\bf k}-{\bf q'}).{\bf r}_{IJ}}
\label{eq:byond-super-cell-self energy-k2}
\end{eqnarray}
Eq.\ref{eq:byond-super-cell-self energy-k2} is centerpiece of our approximation. By iteration, Eq.\ref{eq:byond-super-cell-self energy-k2} up to first order reduces to
\begin{eqnarray}
\Sigma({\bf k}; E)=\frac{1}{N^{2}_{c}}\sum_{IJ}\sum_{{\bf K}_{n}}\Sigma({\bf K}_{n}; E) e^{i({\bf k}-{\bf K}_{n}).{\bf r}_{IJ}}(1-\frac{1}{N_{c}})\nonumber\\
\label{eq:byond-super-cell-self energy-k3}
\end{eqnarray}
where $1<N_{c}<N$. For calculation of self energy $\Sigma({\bf k}; E)$ in Eq.\ref{eq:byond-super-cell-self energy-k3}
first we calculate $\Sigma({\bf K}_{n}; E)$.
 Algorithm for super cell calculation of average Green function is as follows\\
1- A guess is made for real space and K-space self energies,$\Sigma(I,J; E)$, and  $\Sigma({\bf K}_{ n})$. The starting values are usually zero.\\
2- By inserting $\Sigma_{sc}({\bf K}_{ n}; E)$ in Eq.\ref{eq:nth grain green} ,$\bar{G}({\bf K}_{ n};E)=\frac{N_{c}}{N}\sum_{{\bf k}\in\;nth\;grain}(G^{-1}_{0}({\bf k}; E)-\Sigma({\bf K}_{ n};E))^{-1}$,  calculate the grain average k-space Green functions, $\bar{G}({\bf K}_{n}; E)$ .\\
3-From Eq.\ref{eq:math5-7}  calculate K-space cavity Green function $\mathcal{G}({\bf K}_{ n}; E)=({\bar G}^{-1}({\bf K}_{ n};E)+\Sigma({\bf K}_{n}; E))^{-1}$.\\
4- Obtain real space  cavity Green function ${\mathcal G}(I,J; E)=\frac{1}{N_{c}}\sum_{{\bf K}_{n}}e^{i{\bf K}_{n}.{\bf r}_{IJ}}\mathcal{G}({\bf K}_{ n}; E)$ by Fourier transform of k-space $\mathcal{G}({\bf K}_{ n}; E)$.\\
5- Calculate real space super cell impurity Green function matrix $G^{imp}=({\mathcal G}^{-1}-\mathbf{\varepsilon})^{-1}$ .\\
6- Calculate super cell impurity average Green function matrix $\bar{ G}(I,J; E)=<({\mathcal G}^{-1}-\mathbf{\varepsilon})^{-1}>_{IJ}$ by taking average over all possible impurity configurations.\\
7-Calculate real space new super cell self-energy matrix from ${\Sigma}_{sc}={\mathcal G}^{-1}-{\bf \bar G}^{-1}$.\\
8-Inverse Fourier transform of new average super cell self-energy to calculate ${\Sigma}({\bf K}_{n}; E)=\frac{1}{N_{c}}\sum_{IJ}e^{-i{\bf K}_{n}.{\bf r}_{IJ}}{\Sigma}_{sc}(I,J; E)$.\\
9- Return to 2 and repeat until convergence.\\
10- Calculate self-energy beyond super cell approximation by substitution ${\Sigma}({\bf k}; E)$ in Eq.\ref{eq:byond-super-cell-self energy-k3}.\\
11-Calculate average green function from ${\bar G}({\bf k};E)=G^{-1}_{0}({\bf k}; E)-\Sigma({\bf k}; E)$.\\

\begin{figure}[ht]
\centering
\includegraphics[width=7cm]{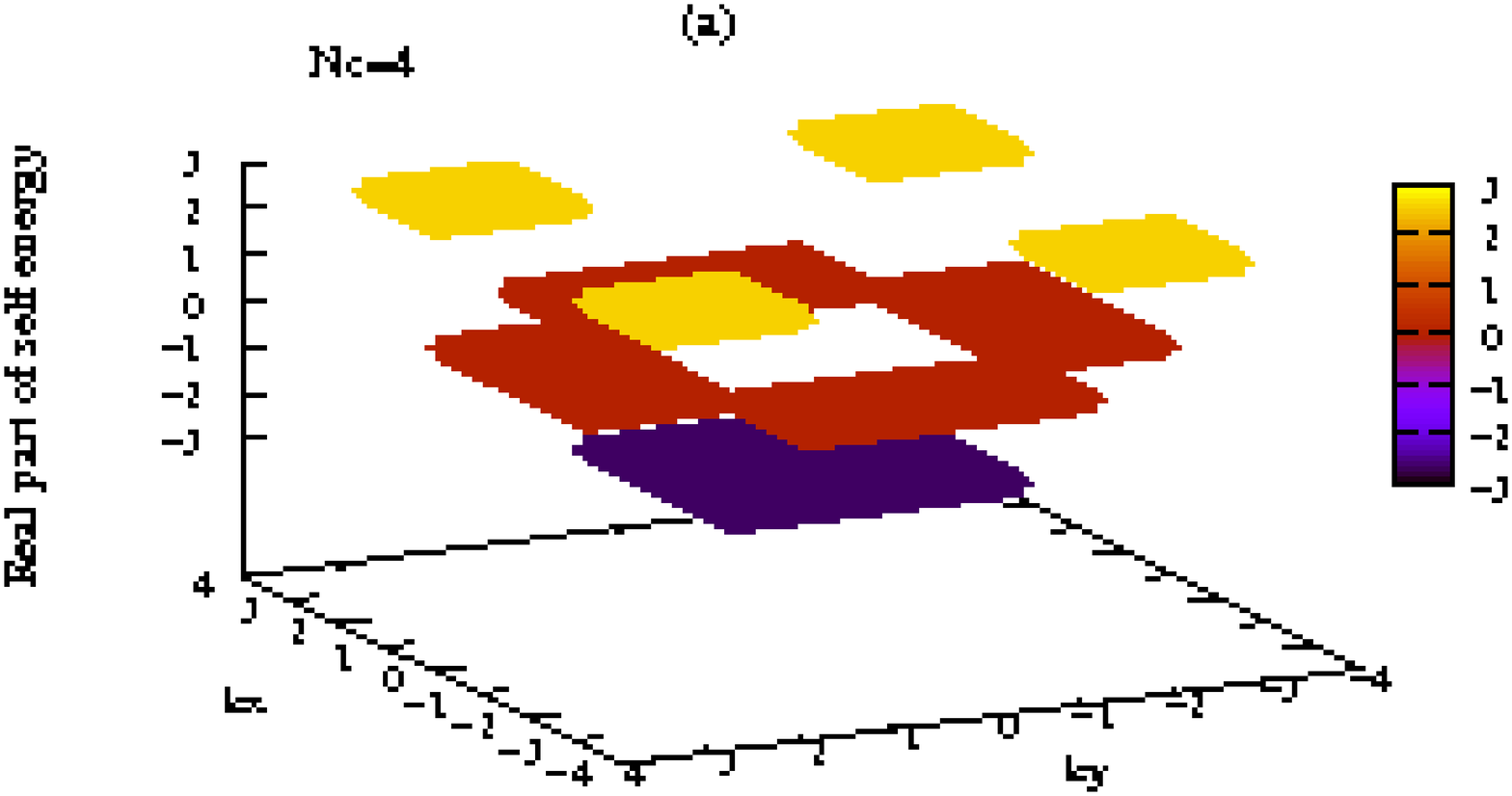}
\includegraphics[width=7cm]{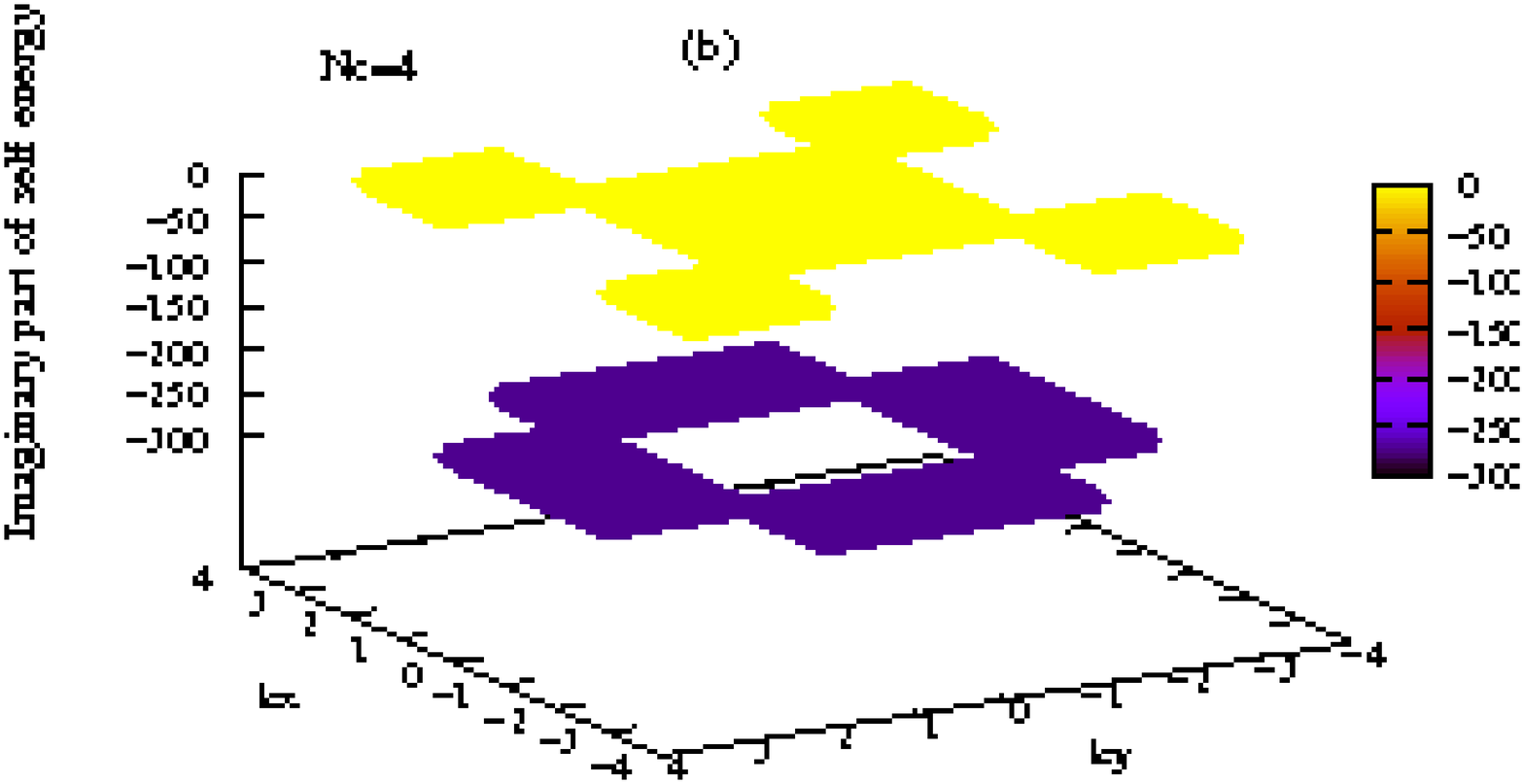}
\includegraphics[width=7cm]{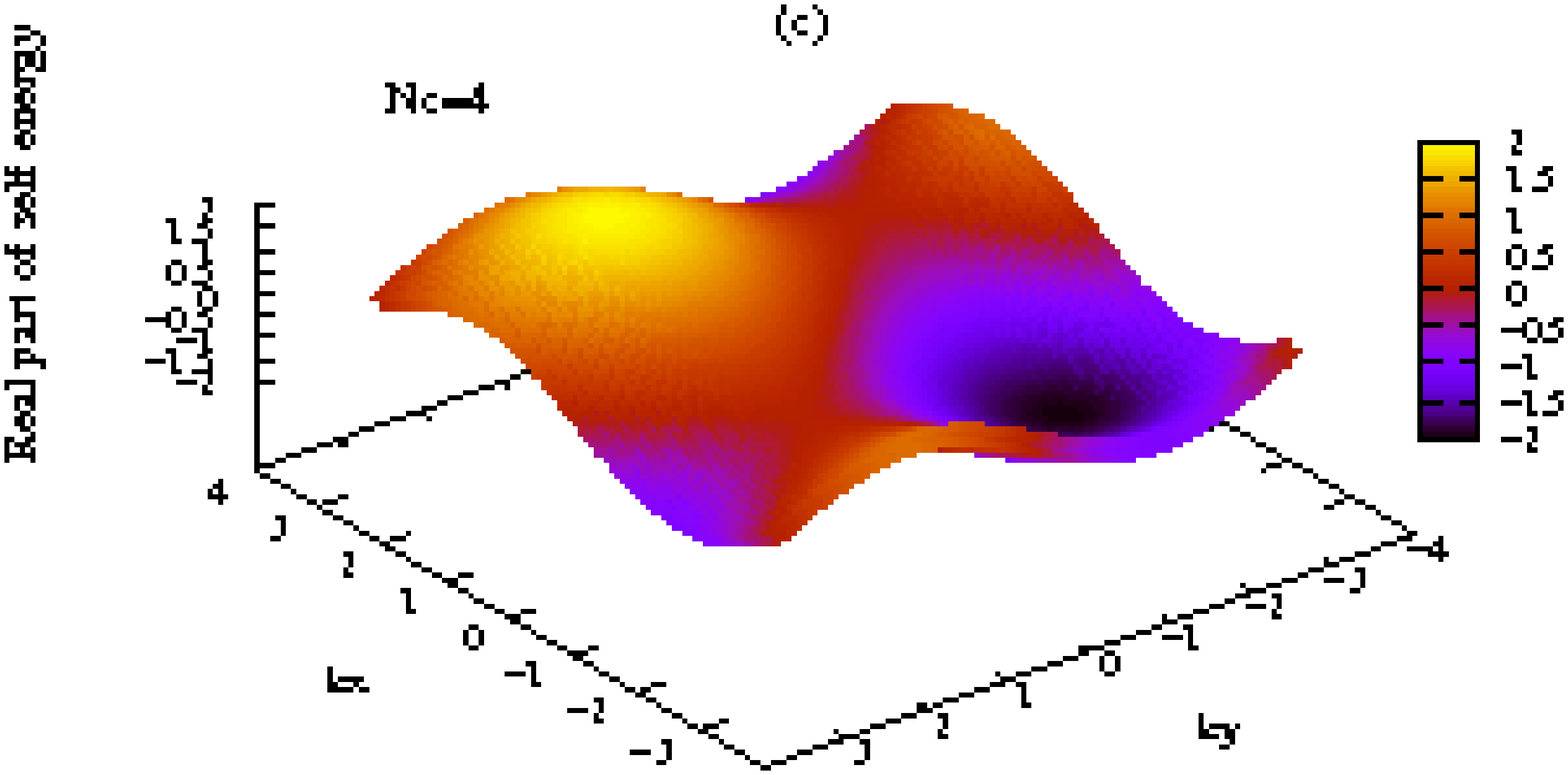}
\includegraphics[width=7cm]{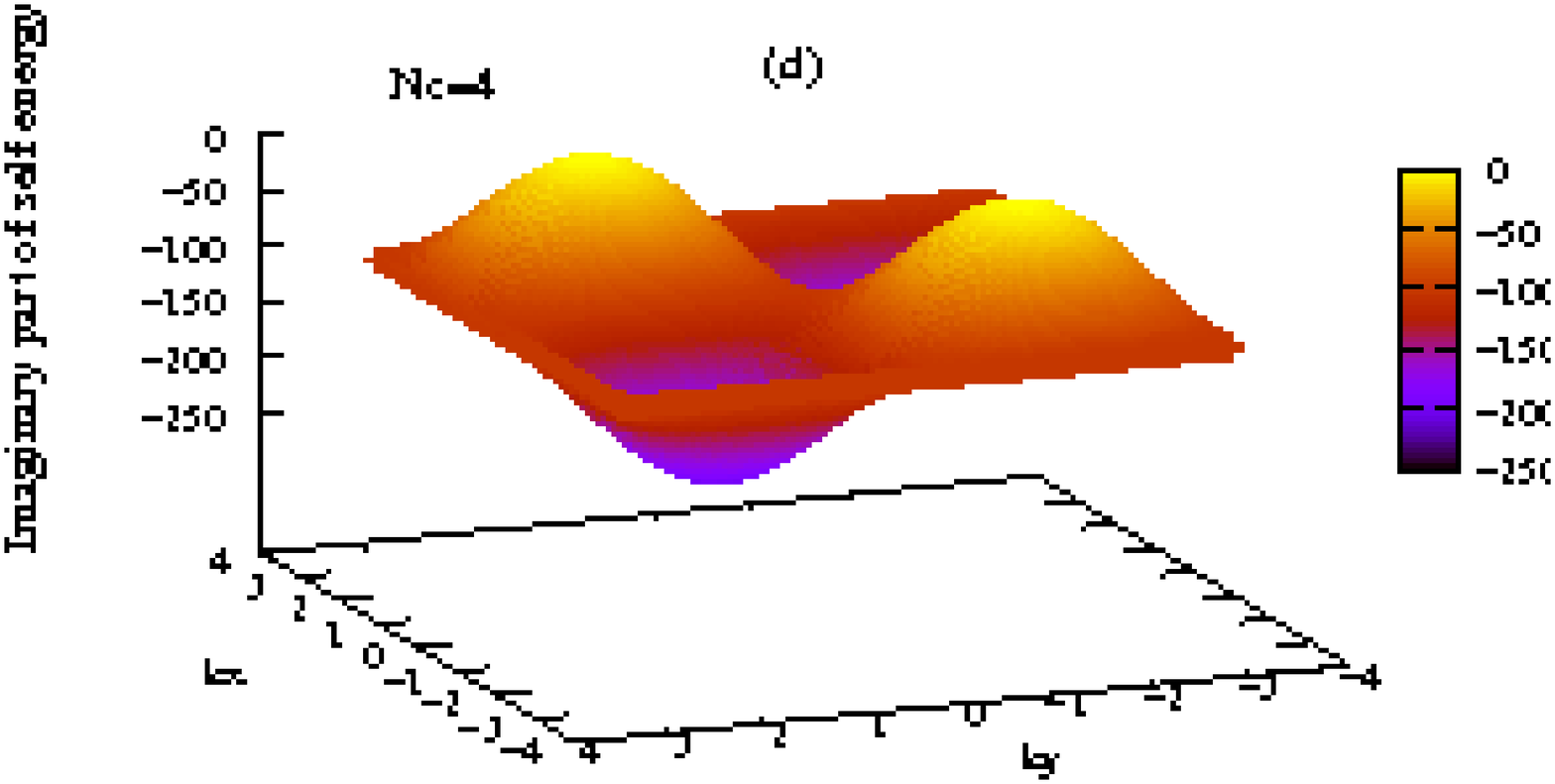}
\caption{(color online)  (a) and (b) show real  and imaginary parts of super cell self energies $\frac{1}{4t}\Sigma({\bf K}_{n},0+i\eta)$ for the $N_{c}=4$ of a two dimensional square alloy. (c) and (d) shows real and imaginary parts of self energy $\frac{1}{4t}\Sigma({\bf k},0+i\eta)$ of a two dimensional square alloy system in beyond $N_{c}=4$ super cell approximation for $\delta=3t$, $c=0.5$ and $\mu=0$. In the super cell approximation k-space self energy in FBZ is discontinuous and k-independent in each grain, but in our beyond super cell approximation it is continuous and fully k-dependent. }
 \label{figure:Re-Im-selfenergy-nc4-d3t-n1-mu0}
\end{figure}

 Now we apply this method to a two dimensional square alloy system in which $\delta=3t$, $c=0.5$ and $\mu=0$.  For this system we calculate, self-energy and density of states in the super cell and beyond super cell approximations and compared them. Fig.\ref{figure:Re-Im-selfenergy-nc4-d3t-n1-mu0} (a) and (b) shows real and imaginary part of self energy$\Sigma({\bf K}_{n};0)$ in terms of  $k_{x}$ and $k_{y}$ in super cell approximation for $N_{c}=4$.   self-energy at the borders of grains have discontinuity and inside of each grain is k-independent. (c) and (d) shows real and imaginary parts of self-energy $\Sigma({\bf k};0)$ in the beyond $N_{c}=4$ super cell approximation which are fully k-dependent and causal.

\begin{figure}[ht]
\centering
\includegraphics[width=5cm]{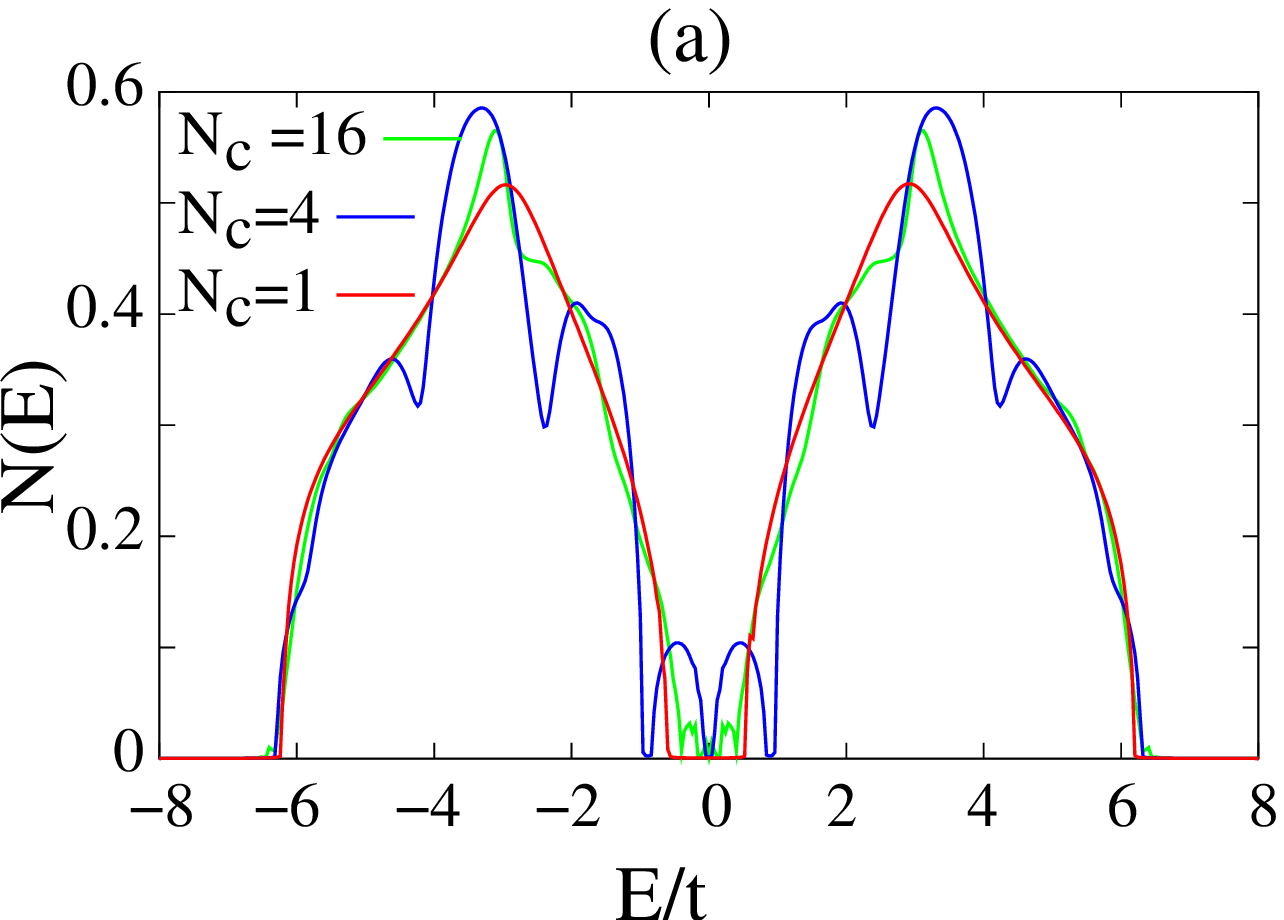}
\includegraphics[width=5cm]{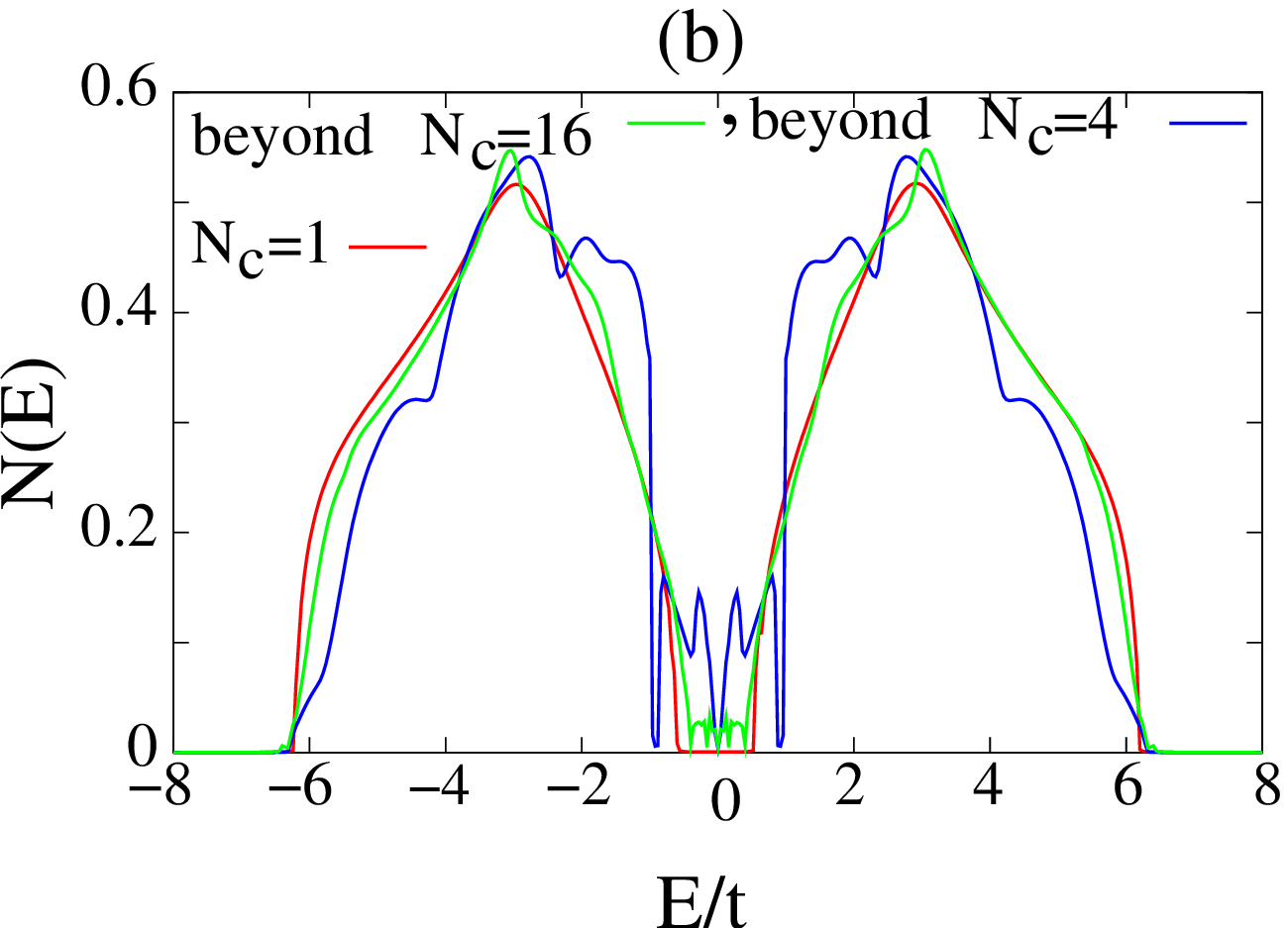}
\caption{(color online) Show comparison of average density of states of a two dimensional square alloy system for (a) CPA  $N_{c}=1$, super cell approximation $N_{c}=4$, $N_{c}=16$ and (b) beyond $N_{c}=4$ and $N_{c}=16$ super cell approximation. The strength length, $\delta=3t$, impurity concentration is $c=0.5$ and $\mu=0$. The difference between density of states are due to nonlocal corrections.}
 \label{figure:nc-nc4-d3t-n1-mu0}
\end{figure}

 Fig.\ref{figure:nc-nc4-d3t-n1-mu0} (a) shows calculated average density of states for super cells $N_{c}=1$, $N_{c}=4$, and $N_{c}=16$ for $\delta=3t$, $c=0.5$ and band filling ${\bar n}=1$. (b) shows average density of states calculated by our beyond super cell approximation for $N_{c}=1$, $N_{c}=4$, and $N_{c}=16$. However bands of this system in this regime splitted in CPA and super cell approximation but in our approximation beyond super cell it is at beginning of splitting.

\begin{figure}[ht]
\centering
\includegraphics[width=6cm]{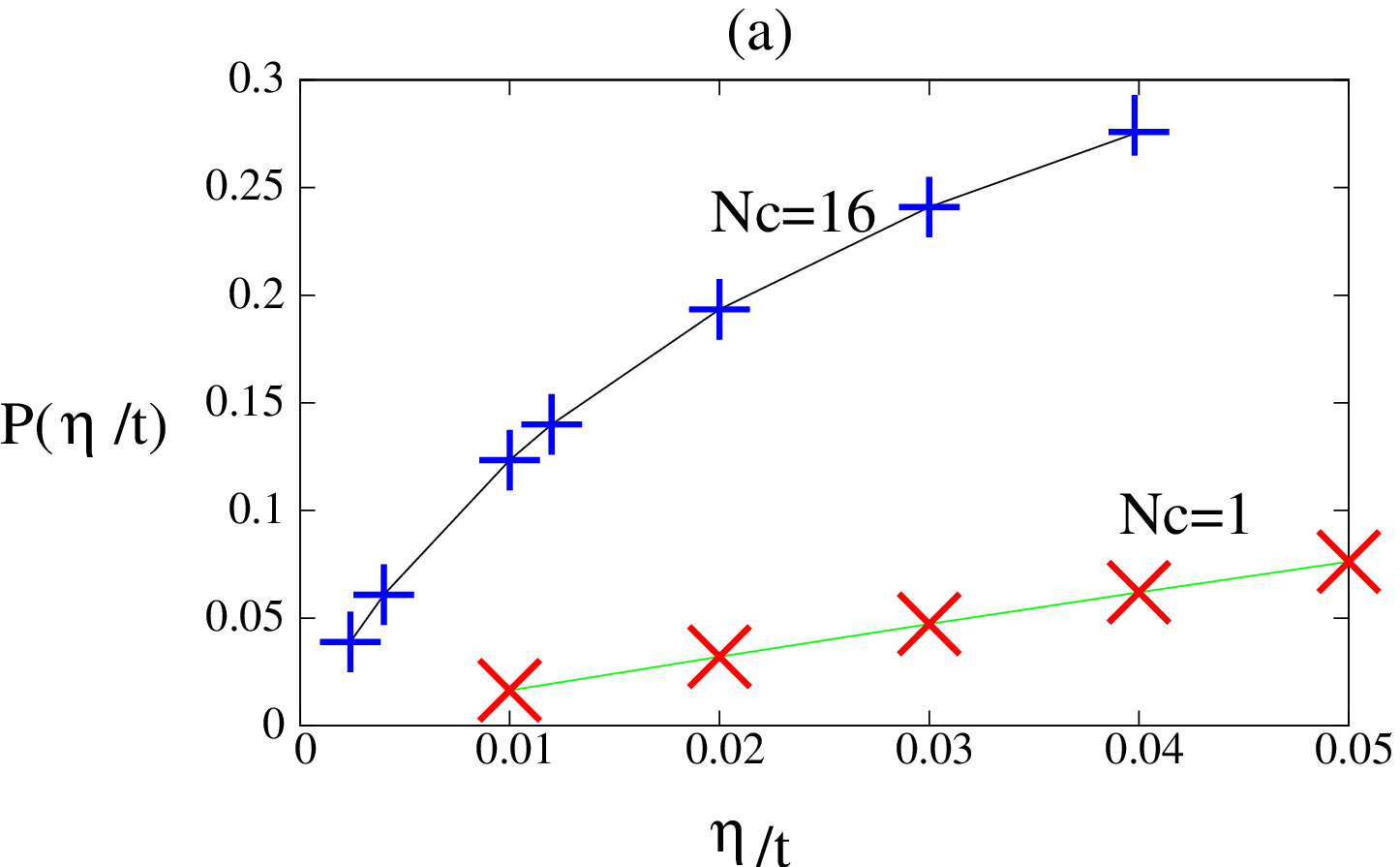}
\includegraphics[width=6cm]{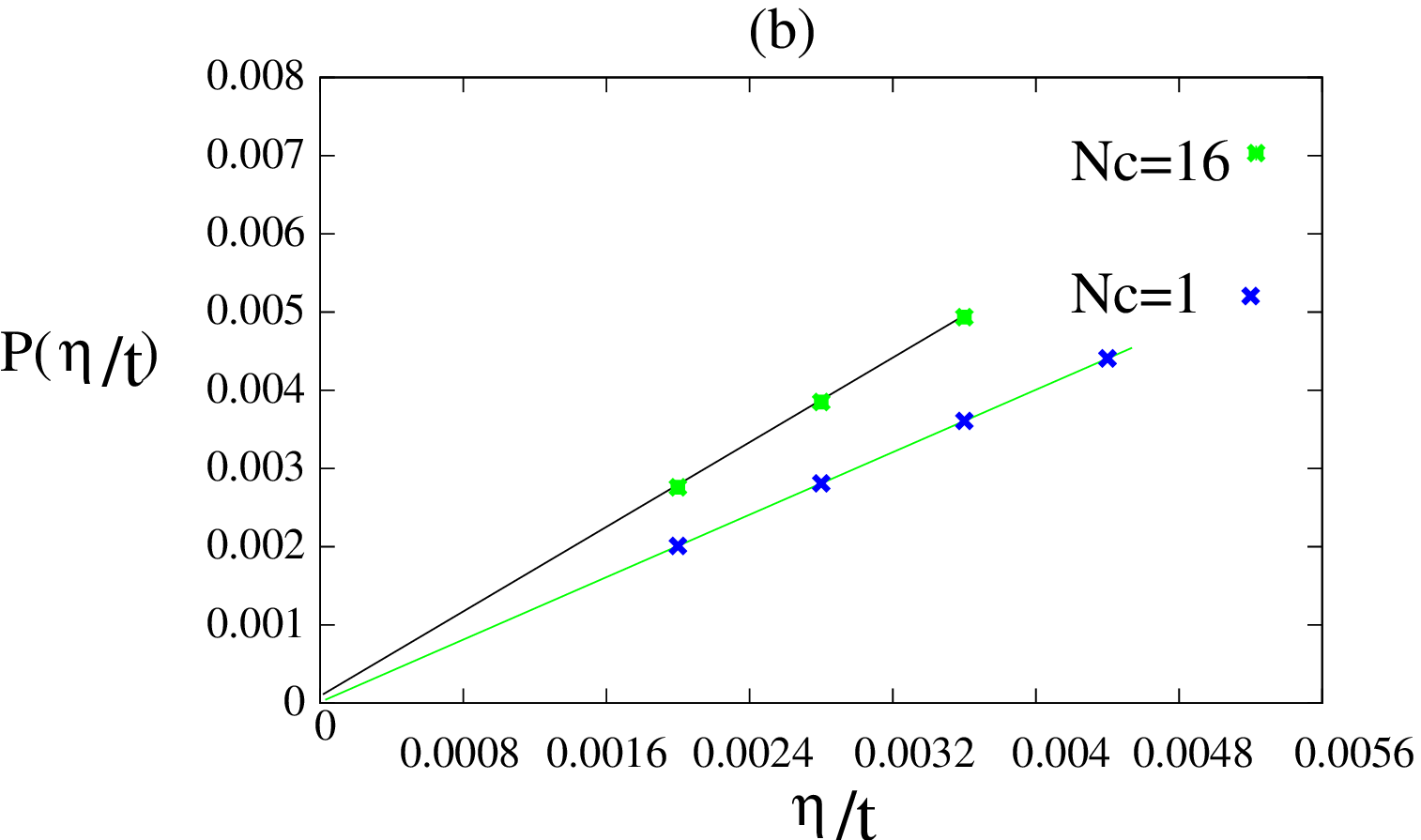}
\caption{(color online) Shows electron localization probability at site $l$ for: (a) one dimensional alloy for $\delta=3t$, $c=0.5$ and average band filling ${\bar n}=1$. For $N_{c}=1$, $P(\frac{\eta}{t})$ extrapolated to zero while for $N_{c}=16$ is fitted to $P(\frac{\eta}{t})=0.007613+14.21 (\frac{\eta}{t})-300.7 (\frac{\eta}{t})^{2}+28.28(\frac{\eta}{t})^{3}$. Hence probability of localization is $P(0)=0.007613$. (b) a two dimensional square alloy system for CPA  $N_{c}=1$ and super cell approximation $N_{c}=16$. The strength length, $\delta=3t$, impurity concentration is $c=0.5$ and $\mu=0$. CPA doesn't shows localization but for $N_{c}=16$ super cell, $P(\frac{\eta}{t}=0)=2.45\times 10^{-5}$ which is due to electron back scattering in the super cell.}
 \label{figure:localization-nc1-nc16}
\end{figure}

One of advantage of super cell approximation is to take in to account electron localization in one and two dimensional disordered alloys which calculates by\cite{T, McK}
\begin{eqnarray}
P(\infty)&=&lim_{t\;\rightarrow\:\infty} <|G_{ll}(E)|^{2}>\nonumber\\&=&lim_{\eta\;\rightarrow\;0}\int d\epsilon  <|G_{ll}(\epsilon+i\eta)|^{2}>
\label{eq:local-super-cell}
\end{eqnarray}
Fig.\ref{figure:localization-nc1-nc16}  shows probability of remaining electron at site $l$ for (a) a one dimensional lattice in the CPA and $N_{c}=16$ super cell approximations for $\delta=3t$, $c=0.5$ and $\mu=0$. CPA $P(\frac{\eta}{t})$ extrapolated to zero while for $N_{c}=16$ it is fitted by  $P(\frac{\eta}{t})=0.007613+14.21 (\frac{\eta}{t})-300.7 (\frac{\eta}{t})^{2}+28.28(\frac{\eta}{t})^{3}$ hence $P(0)=0.007613$. (b) shows it for a square two dimensional alloy in the CPA and
$N_{c}=16$ super cell approximation. In the CPA it is extrapolates to zero but for $N_{c}=16$ it is extrapolating to non zero value $P(\frac{\eta}{t}=0)=2.45\times 10^{-4}$ .

\section{Conclusion}
A successful approximation beyond super cell approximation is introduced. In this approximation self-energy is casual and full k-dependent in the first Brillouin zone. For derivation of the approximation, the entire lattice is divided to super cells with $N_{c}$ sites and no overlap. We proved that self-energy of one site in a definite super cell but another in other super cells are periodic with respect to super cell lengths. Correction to  k-space super cell self-energy comes from sites in different super cells. We added this part to the k-space super cell self-energy. Our approximation recovers CPA in the single site cell limit and as the number of super cell sites approaches the number of lattice sites, $N_{c}\longrightarrow N$, becomes exact. This approximation opens a new channel for observing multi sites scattering effects such as localization that are not observed by other approximations such as DCA and CMDFT. It is overcomes discontinuity and weakly k-dependent of DCA and CMDFT especially for low dimensional systems that k-space self energy is k-dependent significantly. Also it is applicable to both disordered and interacting systems.

\end{document}